\documentclass{jande}
\setVolume{00}{00} 
\setIndex{0000}{00}{000} 
\setPage{1}           

\usepackage[T1]{fontenc}
\usepackage{ae,aecompl}
\usepackage{epstopdf}

\graphicspath{{figures/}}
\allowdisplaybreaks

\theoremstyle{definition}  
\theoremstyle{remark}      
\theoremstyle{plain}       

\theoremstyle{remark}

\theoremstyle{definition}

\titlemark{ Author 1, Author 2 / Journal of Applied Nonlinear Dynamics 00(0) (2022) page1--page2 }
\authormark{Author 1, Author 2 / Journal of Applied Nonlinear Dynamics 00(0) (2022)  page1--page2}
\usepackage{lineno}

\begin{document}

\title{Novel analytical solutions to a new formed model of the (2+1)-dimensional BKP equation using a novel expansion technique}

\author{Rajib Mia \corrauth{rajib.miafma@kiit.ac.in}}

\address{Department of Mathematics, School of Applied Sciences, Kalinga Institute of Industrial Technology (KIIT) Deemed to be University, Bhubaneswar, 751024, Odisha, India}

\Abstract{In this article, we present a comprehensive analytical study to obtain the exact traveling wave solutions to a new formed model of the (2+1)-dimensional BKP equation. We construct exact solutions of the considered model using a  recently developed expansion technique. This current proposed technique has been successfully implemented to obtain a few exact solutions of a new formed (2+1)-dimensional  BKP equation.  In order to understand the physical interpretation of  solutions effectively,  the 2D and 3D graphs are  plotted for each type of the solutions obtained  for different particular values of the parameters. Furthermore, it is found that  the obtained solutions are periodic and solitary wave solutions.
We  anticipate that the proposed method is reliable and can be applied for obtaining wave solutions of  the other nonlinear evolution equations (NLEEs).}[\hfill Nonlinear PDEs\par \hfill BKP equation\par \hfill Exact solution\par \hfill Traveling wave solutions
]

\maketitle

\thispagestyle{firstpage}
\renewcommand{\baselinestretch}{1}
\normalsize


\section{Introduction}
Nowadays the exact solution of nonlinear evolution equation is a very active research area. In the study of nonlinear physical phenomena, the exact analytical traveling wave solutions of the non--linear partial--differential equations (NPDEs)  plays a very significant role. The non-linear physical wave phenomena takes place in several engineering and scientific fields. For example, in  plasma physics, optical fibers, chemical physics and geochemistry, biology, solid state physics and  fluid mechanics,  the non-linear evolution equations are broadly used as a model to study the  physical phenomena.\\

Various powerful effective techniques have been established to obtain the exact traveling wave solutions of various nonlinear evolution equations. Some of these powerful techniques are the sine-Gordon method \cite{ali2019analytical,akbar2021soliton,fahim2022wave},  the Sardar
sub-equation technique \cite{rezazadeh2020new,cinar2022derivation,asjad2022traveling}, the modified Sardar sub-equation method \cite{ur2022solitary,ozdemir2022novel}, the Riccati-Bernoulli sub-ODE method \cite{yang2015riccati,hassan2019riccati}, the $\left( \frac{G'}{G}\right)$-expansion method \cite{wang2008g,zayed2009some,akbar2012abundant,kim2012new,mamun2021solitary,aniqa2022soliton}, 
the $\left( \frac{G'}{G}, \frac{1}{G}\right)$-expansion method \cite{li2010g,zayed2012traveling,miah2020abundant,yokus2020construction},
the modified exp-function method \cite{naher2012new,biazar2012exp,shakeel2022application,yi2021modified}, 
the simple equation method (SEM) \cite{zayed2011note,mirzazadeh2014modified}, 
the tanh--coth expansion method \cite{parkes2010observations,manafian2016comparison,ali2022variety}, $\left(\frac{1}{G'}\right)$-expansion technique \cite{yokucs2019complex,gholami2020extraction,yokus2020construction,durur2021exact}, Lie symmetry approach \cite{khalique2009lie,kumar2018soliton,khalique2020coupled,kumar2021application} and so on.

Recently, Ahmed et al.  have found mixed lump, periodic-lump and breather soliton solutions  to the (2 + 1)-dimensional KP equation with the help of symbolic computation using the Hirota bilinear method \cite{ahmed2019mixed}. They have also observed some new type of characteristics
of the periodic--lump solutions and kinky--breather solitons. 
Kayum et al.  have established stable traveling wave solutions to the nonlinear Klein--Gordon equation in particle physics, condensed matter physics,   solid state physics, the gas dynamics and nonlinear optics equation utilizing the sine-Gordon expansion technique \cite{kayum2021onset}. They have established  the familiar stable wave solutions covering an extensive range which related to free parameters. They have obtain various new solutions for the particular values of the parameters.
In the article \cite{roy2021bright},  a generalized ($G'/G$)-expansion technique is used for deriving the  exact solitary wave solutions generated in the form of  trigonometric, hyperbolic and rational functions for the the non-linear and regularized long wave (RLW) as well as for the Riemann--wave (RW) models.
Shakeel et al. \cite{shakeel2022application} have obtained various types of analytical exact wave solutions such as  complex function  and hyperbolic solutions of the strain-wave equation found in micro-structured solids, which is very important in the field of  solid physics, utilizing the modified exp-function technique.

Tripathy et al. \cite{tripathy2020exact} and Khaliq et al. \cite{khaliq2022some} have successfully applied the proposed expansion technique to obtain  some novel exact traveling wave solutions of  the ion sound Langmuir wave model  and  the (2 + 1)-dimensional Boussinesq equation respectively. They have demonstrated  the physical interpretation of nonlinear processes and the efficiency of the proposed method. 



In this current work, our main purpose is to obtain the new exact traveling wave solutions to a new formed equation of  the $(2+1)$-dimensional  BKP model. In this work, for the first time we have used  $\left(\frac{G'}{G'+G+A}\right)$ -expansion technique \cite{hong2019g,khaliq2022some,mia2023new} to obtain a few analytical solutions to a  new formed equation of  the $(2+1)$-dimensional BKP model. We have plotted the 3D and 2D graphics simulations of the newly obtained exact solutions. By observing the various nature of the waves and from numerical simulations one can find more useful information regarding the new reduced form of  the $(2+1)$-dimensional generalized BKP equation.

This work is  arranged  as follows: In the Section \ref{sec:2}, we have explained the algorithm of the $\left( \frac{G'}{G'+G+A}\right)$- expansion technique.  In Section \ref{sec:3}, we have implemented the proposed method to obtain the exact traveling wave solutions to a new form of  the $(2+1)$-dimensional B-type Kadomtsev-Petviashvili (BKP) equation.  Results and discussions are given in Section \ref{sec:4}. Lastly, Section \ref{sec:5} is devoted to the conclusions.

\section{The algorithm of the $\left( \frac{G'}{G'+G+A}\right)$- expansion technique} \label{sec:2}
The  steps of the $\left( \frac{G'}{G'+G+A}\right)$-expansion technique to find the exact traveling wave solutions of the nonlinear evolution equations (NLEEs)  are discussed in this section. 
Consider the general non-linear evolution equation of the following type
\begin{equation}
f(u, u_y, u_x, u_t, u_{xx}, u_{yy}, u_{yx}, u_{xt}, u_{xy}, \dots)=0, \label{eq:2}
\end{equation}
where the function $u\equiv u(x, y, t)$ is not known, $f$ is a polynomial function involving  $u$ and its higher order derivatives together with  nonlinear terms. Subscripts in this Eq. (\ref{eq:2})  indicate the various partial derivatives.
The main steps of the $\left( \frac{G'}{G'+G+A}\right)$- expansion technique are  given below:\\
\textbf{Step-I}: The traveling wave variable is defined as $\xi=n x+m y-p t$ by taking three independent variables $x ,y$, and $t$ into one variable $\xi$. Here $n$, $m$ and $p$ are constants. 
Let us assume that the solution of Eq. (\ref{eq:2}) in $\xi$ as 
\begin{eqnarray}
u(x, y, t)=w(\xi). \label{eq:3}
\end{eqnarray}
Using the above transformation  $\xi=n x+m y-p t$ and $u(x, y, t)=w(\xi)$ in the Eq. (\ref{eq:2}), we get the following ordinary differential equation (ODE):
\begin{eqnarray}
g(w, w', w'', w''', \dots)=0, \label{eq:4}
\end{eqnarray}
where $w'=\frac{dw}{d\xi},~w''=\frac{d^2w}{d\xi^2},~w'''=\frac{d^3 w}{d\xi^3}\dots$.\\
\textbf{Step-II}:
We suppose that the exact analytical solutions of  Eq. (\ref{eq:4}) can be written in the following form:
\begin{eqnarray}
w(\xi)=\sum_{k=0}^{N} a_i \left(\frac{G'}{G'+G+A}\right)^{k}, \label{eq:5}
\end{eqnarray} 
where $G\equiv G(\xi)$ satisfies the following auxiliary second order ODE:
\begin{eqnarray}
G''+B G'+C G+A C=0. \label{eq:6}
\end{eqnarray}
Here the prime indicates for ordinary derivative with respect to $\xi$. In the above equation $A, B, C$ are real constants and $a_k (k=0, 1, 2, \dots, N)$ are arbitrary constants.\\
\textbf{Step-III}:
By the homogeneous balance principle and balancing the  highest order derivative of $w$ with the highest order nonlinear term in Eq. (\ref{eq:4}), determine the positive integer $N$.
Furthermore, the coefficients $a_k~(k=0, 1, 2, \dots N )$ can be found by solving a system of linear algebraic equations which will come from suggested method. Then using the values of $A, B$ and $C$ and  Eq. (\ref{eq:6}), the function $G(\xi)$ will be evaluated.\\
\textbf{Step-IV}:
Lastly, on substitution of  Eq. (\ref{eq:5}) in the Eq. (\ref{eq:4}), we can obtain the polynomial of  $\left( \frac{G'}{G'+G+A}\right)$. Then by equating  the  coefficients of like powers of  $\left( \frac{G'}{G'+G+A}\right)$ from both sides, we will get a system of algebraic equations for $\eta$, $a_k~(k=0, 1, 2, \dots N )$, $A, B,$ and $C$.

\section{Implementation of the proposed technique} \label{sec:3}
In this current section, we have presented the implementation  of the proposed  $\left( \frac{G'}{G'+G+A}\right)$-expansion method to derive the exact analytical  wave solutions to a new reduced form of  the $(2+1)$-dimensional BKP model. The reduced new form $(2+1)$-dimensional  BKP model could be obtained  from the (3+1)-dimensional generalized  BKP model by setting the spatial variable $z=x$ \cite{kaur2018lump}.  The new formed equation of the  $(2+1)$-dimensional  BKP model is presented below \cite{kaur2018lump,kara2022analytical}:
\begin{eqnarray}
u_{xxxy}+\alpha (u_{y} u_{x})_x+(u_y+2u_x)_t -(u_{yy}+2u_{xx})=0, \label{eq:1}
\end{eqnarray}
where $\alpha$ is a constant. Recently Kara et al. \cite{kara2022analytical} applied the $\left(\frac{G'}{G}, \frac{1}{G}\right)$-expansion method to the new form of the $(2+1)-$dimensional BKP equation and obtained exact solutions in the form of hyperbolic, trigonometric and rational functions.
Utilizing the wave  transformation $u(x, y, t)=w(\xi)$ where $\xi=n x+m y-p t$, the Eq. (\ref{eq:1}) reduces to the following ODE
\begin{eqnarray}
n^3 m w^{iv}+2\alpha n^2 m w' w''-(2n^2+m^2+2 n p+m p)w''=0.\label{eq:7}
\end{eqnarray}
By assuming $w'(\xi)=U(\xi)$ and $\eta=-(2 n^2+m^2+2np+mp)$,  the Eq. (\ref{eq:7}) reduces to
\begin{eqnarray}
n^3 m U'''+2\alpha n^2 m U U'+\eta U'=0. \label{eq:8}
\end{eqnarray}
Integrating  Eq. (\ref{eq:8}) with respect to $\xi$ and assuming the integration constant equal to
zero, we obtain
\begin{eqnarray}
n^3 m U''+ \alpha n^2 m U^2+\eta U=0. \label{eq:9}
\end{eqnarray} 
By the homogeneous balance principle, balancing between the terms $U''$ and $U^2$ in Eq. (\ref{eq:9}), gives  $N+2=2N$ which implies $N=2$.
Hence from Eq. (\ref{eq:5}),  the solution of Eq. (\ref{eq:9}) can be written as:
\begin{eqnarray}
U(\xi)=\sum_{i=0}^{2} a_i \left(\frac{G'}{G'+G+A}\right)^{i}. \label{eq:10}
\end{eqnarray} 
Utilizing Eq. (\ref{eq:10}) into Eq. (\ref{eq:9}) and equating the coefficients of  similar powers of $\left(\frac{G'}{G'+G+A}\right)$ of 
Eq. (\ref{eq:9}), we obtain a system of linear algebraic equations which is given below:
\begin{eqnarray}
\begin{cases}
a_1 B C m n^3-2 a_1 C^2 m n^3+2 a_2 C^2 m n^3+a_0 \eta +\alpha  a_0^2 m n^2=0,\\
a_1 B^2 m n^3-6 a_1 B C m n^3+6 a_2 B C m n^3+6 a_1 C^2 m n^3\\
~~~~~-12 a_2 C^2 m n^3+2 a_1 C m n^3+a_1 \eta +2 \alpha  a_0 a_1 m n^2=0,\\\label{eq:1eq11}
-3 a_1 B^2 m n^3+4 a_2 B^2 m n^3+9 a_1 B C m n^3-24 a_2 B C m n^3\\
~~~~~+3 a_1 B m n^3-6 a_1 C^2 m n^3+24 a_2 C^2 m n^3-6 a_1 C m n^3\\
~~~~~+8 a_2 C m n^3+a_2 \eta +\alpha  a_1^2 m n^2+2 \alpha  a_0 a_2 m n^2=0,\\
2 a_1 B^2 m n^3-10 a_2 B^2 m n^3-4 a_1 B C m n^3+30 a_2 B C m n^3\\
~~~~~-4 a_1 B m n^3+10 a_2 B m n^3+2 a_1 C^2 m n^3-20 a_2 C^2 m n^3\\
~~~~~+4 a_1 C m n^3-20 a_2 C m n^3+2 a_1 m n^3+2 \alpha  a_1 a_2 m n^2=0,\\
6 a_2 B^2 m n^3-12 a_2 B C m n^3-12 a_2 B m n^3+6 a_2 C^2 m n^3\\
~~~~~+12 a_2 C m n^3+6 a_2 m n^3+\alpha  a_2^2 m n^2=0.
\end{cases}
\end{eqnarray} 
By solving the above system in Eq. (\ref{eq:1eq11}), we obtain two sets of solutions as given below:\\
\textbf{\textit{SET-1}}:
\begin{eqnarray}
&&\eta =4 C m n^3-B^2 m n^3, ~~a_0=-\frac{6 C n (-B+C+1)}{\alpha }, \nonumber\\
&&a_1= \frac{6 \left(B^2 n-3 B C n-B n+2 C^2 n+2 C n\right)}{\alpha },~
a_2=-\frac{6 n (B-C-1)^2}{\alpha }\nonumber
\end{eqnarray}
\textbf{\textit{SET-2}}:
\begin{eqnarray}
&&\eta = m n^3 \left(B^2-4 C\right),~~a_0=\frac{-B^2 n+6 B C n-6 C^2 n-2 C n}{\alpha },\nonumber\\
&&a_1= \frac{6 \left(B^2 n-3 B C n-B n+2 C^2 n+2 C n\right)}{\alpha },~
a_2=-\frac{6 n (B-C-1)^2}{\alpha }\nonumber
\end{eqnarray}
For \textbf{\textit{SET-1}}, we obtain the following exact traveling wave solutions:\\
Case-I: When $\Lambda=B^2-4C>0$
\begin{eqnarray}\label{eq:16}
&&U_{11}(x, y, t)= -\frac{6 n C  (C-B+1)}{\alpha } \nonumber\\
&&+\frac{6 \left(B^2 n-3 B C n-B n+2 C^2 n+2 C n\right)}{\alpha }
\left[ \frac{C_1 \left(B+\sqrt{\Lambda}\right)+C_2 \left(B-\sqrt{\Lambda}\right) e^{\sqrt{\Lambda} \xi }}{C_1 \left(B+\sqrt{\Lambda}-2\right)+C_2 \left(B-\sqrt{\Lambda}-2\right) e^{\sqrt{\Lambda} \xi }}\right] \nonumber\\
&&-\frac{6 n (B-C-1)^2}{\alpha }
\left[ \frac{C_1 \left(B+\sqrt{\Lambda}\right)+C_2 \left(B-\sqrt{\Lambda}\right) e^{\sqrt{\Lambda} \xi }}{C_1 \left(B+\sqrt{\Lambda}-2\right)+C_2 \left(B-\sqrt{\Lambda}-2\right) e^{\sqrt{\Lambda} \xi }}\right] ^2.
\end{eqnarray}
Case-II: When $\Lambda=B^2-4C<0$
\begin{eqnarray}
&&U_{12}(x, y, t)= -\frac{6 C n (-B+C+1)}{\alpha }\nonumber\label{eq:17}\\
&&+\frac{6 \left(n B^2 -3 n B C -n B +2n C^2 +2 n C \right)}{\alpha }\times\\
&&\left[ \frac{\sin \left(\frac{\sqrt{-\Lambda} }{2}\xi \right) \left(B C_2+C_1 \sqrt{-\Lambda}\right)+\cos \left(\frac{\sqrt{-\Lambda}  }{2}\xi\right) \left(B C_1-C_2 \sqrt{-\Lambda}\right)}{\sin \left(\frac{\sqrt{-\Lambda} }{2} \xi \right) \left((B-2) C_2+C_1 \sqrt{-\Lambda}\right)+\cos \left(\frac{\sqrt{-\Lambda} }{2} \xi\right) \left((B-2) C_1-C_2 \sqrt{-\Lambda}\right)}\right] \nonumber\\
&& -\frac{6 n (B-C-1)^2}{\alpha }\times\nonumber\\
&&\left[ \frac{\sin \left(\frac{\sqrt{-\Lambda} }{2} \xi\right) \left(B C_2+C_1 \sqrt{-\Lambda}\right)+\cos \left(\frac{\sqrt{-\Lambda}}{2} \xi \right) \left(B C_1-C_2 \sqrt{-\Lambda}\right)}{\sin \left(\frac{\sqrt{-\Lambda} }{2} \xi\right) \left((B-2) C_2+C_1 \sqrt{-\Lambda}\right)+\cos \left(\frac{\sqrt{-\Lambda} }{2}\xi \right) \left((B-2) C_1-C_2 \sqrt{-\Lambda}\right)}\right]^2\nonumber.
\end{eqnarray}
For \textbf{\textit{SET-2}},  we get the following exact traveling wave solutions:\\
Case-I: When $\Lambda=B^2-4C>0$
\begin{eqnarray}\label{eq:18}
&&U_{21}(x, y, t)=  \frac{-B^2 n+6 B C n-6 C^2 n-2 C n}{\alpha }\nonumber\\
&&+~~~ \frac{6 \left(B^2 n-3 B C n-B n+2 C^2 n+2 C n\right)}{\alpha }
\left[ \frac{C_1 \left(B+\sqrt{\Lambda}\right)+C_2 \left(B-\sqrt{\Lambda}\right) e^{\sqrt{\Lambda} \xi }}{C_1 \left(B+\sqrt{\Lambda}-2\right)+C_2 \left(B-\sqrt{\Lambda}-2\right) e^{\sqrt{\Lambda} \xi }}\right] \nonumber\\
&& -\frac{6 n (B-C-1)^2}{\alpha }
\left[ \frac{C_1 \left(B+\sqrt{\Lambda}\right)+C_2 \left(B-\sqrt{\Lambda}\right) e^{\sqrt{\Lambda} \xi }}{C_1 \left(B+\sqrt{\Lambda}-2\right)+C_2 \left(B-\sqrt{\Lambda}-2\right) e^{\sqrt{\Lambda} \xi }}\right] ^2\nonumber.
\end{eqnarray}

Case-II: When $\Lambda=B^2-4C<0$
\begin{eqnarray}
&&U_{22}(x, y, t)=   \frac{-B^2 n+6 B C n-6 C^2 n-2 C n}{\alpha } + \frac{6 \left(B^2 n-3 B C n-B n+2 C^2 n+2 C n\right)}{\alpha }\times\label{eq:19}\\
&&\left[ \frac{\sin \left(\frac{\sqrt{-\Lambda}}{2} \xi \right) \left(B C_2+C_1 \sqrt{-\Lambda}\right)+\cos \left(\frac{\sqrt{-\Lambda}  }{2}\xi\right) \left(B C_1-C_2 \sqrt{-\Lambda}\right)}{\sin \left(\frac{\sqrt{-\Lambda} }{2}\xi \right) \left((B-2) C_2+C_1 \sqrt{-\Lambda}\right)+\cos \left(\frac{\sqrt{-\Lambda}  }{2}\xi\right) \left((B-2) C_1-C_2 \sqrt{-\Lambda}\right)}\right]\nonumber \\
&& -\frac{6 n (B-C-1)^2}{\alpha }
\left[ \frac{\sin \left(\frac{\sqrt{-\Lambda} }{2} \xi\right) \left(B C_2+C_1 \sqrt{-\Lambda}\right)+\cos \left(\frac{\sqrt{-\Lambda}  }{2}\xi\right) \left(B C_1-C_2 \sqrt{-\Lambda}\right)}{\sin \left(\frac{\sqrt{-\Lambda}  }{2}\xi\right) \left((B-2) C_2+C_1 \sqrt{-\Lambda}\right)+\cos \left(\frac{\sqrt{-\Lambda}  }{2}\xi\right) \left((B-2) C_1-C_2 \sqrt{-\Lambda}\right)}\right]^2\nonumber.
\end{eqnarray}

\section{Results and discussion} \label{sec:4}

In this section, we outline a variety of traveling wave solutions  representations of the obtained results for the different values of the associated parameters. Integrating $U_{11}$, $U_{12}$, $U_{21}$ and $U_{22}$ given in Eqs. (\ref{eq:16})-(\ref{eq:19}) with respect to $\xi$, we get the required solutions and  have shown the 2D  and 3D plots of four different solutions in different cases (see Figs. \ref{fig:1}-\ref{fig:4}). The kink shape soliton solution for \textbf{\textit{SET-1}}(Case-1) is presented in  Fig. \ref{fig:1}  for the associated parameter values  ``$t = 1; \alpha = 1; B = 1; C = 0.1;   C_1 =1, C_2 = 1$, $n =1 ; m = 1; p = -0.8$". 
In this figure, (I) represents the 3D plot of Eq. (\ref{eq:16}) and curve (II) represents the corresponding 2D plot of Eq. (\ref{eq:16}). 
Fig. {\ref{fig:2} represents the 3D and 2D plots of the singular periodic shape wave solution at the parameter values  ``$t = 1,   \alpha = 1, B = 1,  C = 1.1,   C_1 =1, C_2 = 1$,  $n = 1, m = 1,  p = -2.13333$'' for \textbf{\textit{SET-1}}(Case-2) of Eq. (\ref{eq:17}). 
One soliton solution of Eq. (\ref{eq:18}) for \textbf{\textit{SET-2}}(Case-1) at the parameter values ``$t = 1,  \alpha = 1, B =1,  C = 0.15,   C_1 =1, C_2 = 1$,	$n = 1,  m = 1,  p = -1.13333$'' is depicted in Fig. \ref{fig:3} in which (I) represents the 3D plot and curve (II) represents the corresponding 2D plot of Eq. (\ref{eq:18}). 	In Fig. \ref{fig:4},  we have shown the 3D and 2D graphical representations of the singular periodic wave solution of Eq. (\ref{eq:19}) for \textbf{\textit{SET-2}}(Case-2) for the parameter values ``$t = 1,   \alpha = 1, B =1,  C = 1.1,  C_1 =0, C_2 = 1$, $n = 1, m = 1,  p =0.133333$''.
\begin{figure}[t!]
	\centering
	\includegraphics[height=0.35\textwidth]{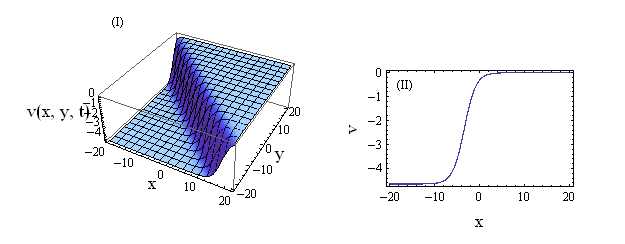}
	\caption{The king shape soliton solution for  $t = 1; B = 1; \alpha = 1; C = 0.1;  C_1 =1, C_2 = 1$: (I) represents the 3D plot  and (II) represents the 2D plot of Eq. (\ref{eq:16}).
		\label{fig:1}}
\end{figure}

\begin{figure}[t!]
	\centering
	\includegraphics[height=0.35\textwidth]{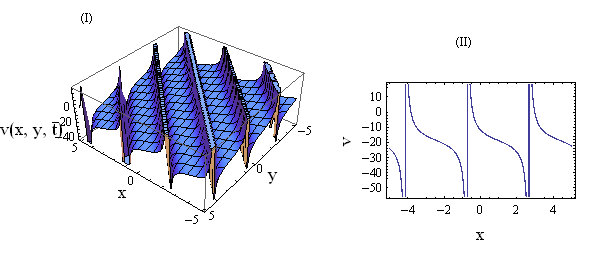}
	\caption{The singular periodic wave solution for  $t = 1; \alpha = 1;  B = 1; C = 1.1; C_1 =1, C_2 = 1$: (I) represents the 3D plot  and (II) represents the 2D plot of Eq. (\ref{eq:17}).}\label{fig:2}
\end{figure}
\begin{figure}[t!]
	\centering
	\includegraphics[height=0.35\textwidth]{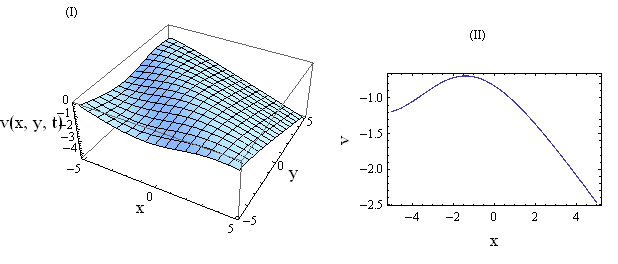}
	\caption{ One soliton wave solution for $t = 1,  B =1,   \alpha = 1, C = 0.15,    C_1 =1, C_2 = 1$: (I) represents the 3D plot  and (II) represents the 2D plot of Eq. (\ref{eq:18})}.\label{fig:3}
\end{figure}

\begin{figure}[t!]
	\centering
	\includegraphics[height=0.35\textwidth]{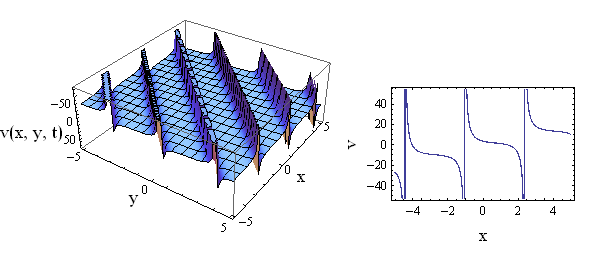}
	\caption{ The singular periodic wave solution for $t = 1, B =1, C = 1.1,  \alpha = 1,   C_1 =0, C_2 = 1$: (I) represents the 3D plot  and (II) represents the 2D plot of Eq. (\ref{eq:19})}.\label{fig:4}
\end{figure}

\section{Conclusions} \label{sec:5}
The relatively  new $\left(\frac{G'}{G'+G+A}\right)$-expansion technique has been successfully utilized to derive the new exact traveling wave solutions to a  new reduced form of  the $(2+1)$-dimensional  BKP equation. We have used the symbolic mathematical computation program (Mathematica), to show the graphical representation of the solutions. The exact traveling wave solutions are presented in the form of 2-D and 3-D figures. It is noticed that all the solutions obtained are in general form and involving a few parameters. The reported solutions are represented by equations (\ref{eq:16}), (\ref{eq:17}), (\ref{eq:18}) and (\ref{eq:19}). By assigning the particular values to the parameters, one kink shape soliton solution, two singular periodic shape wave solutions and one soliton solution are presented in  Figs. \ref{fig:1} -- \ref{fig:4}.

Exploring the new  exact analytical solutions of the new reduced form of  the $(2+1)$-dimensional generalized BKP equation, we anticipate that the aforementioned technique could be applied to many other NPDES model arising in mathematical physics and engineering problems to obtain new exact wave solutions.

\section*{Acknowledgments}
The author would like to thank the anonymous referees and editor for all the apt suggestions and comments which improved both the quality and the clarity of the paper.

\end{document}